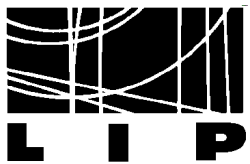

LABORATÓRIO DE INSTRUMENTAÇÃO E
FÍSICA EXPERIMENTAL DE PARTÍCULAS

# Gain, Rate and Position Resolution Limits of Micropattern Gaseous Detectors

V.Peskov[1,*], P.Fonte[2,3]

1 - Royal Institute of Technology, Stockholm, Sweden
2 - CERN, Geneva, Switzerland
3 - LIP/ISEC, Coimbra, Portugal

## Abstract

In this study we report the results of a systematic study of the gain, rate and the position resolution limits of various micropattern gaseous detectors. It was found that at low rates (<1 Hz/mm$^2$) each detector has it own gain limit, which depends on the size and design features, as well as on gas composition and pressure. However, in all cases the maximum achievable gain is less than or equal to the classical Raether limit. It also was found that for all detectors tested the maximum achievable gain drops sharply with the counting rate.

The position resolution of micropattern detectors for detection of X-rays (6 to 35 kV) was also studied, being demonstrated that with solid converters one could reach a position resolution better than 30 µm at 1 atm in a simple counting mode.

---

[*] Corresponding author: V. Peskov, Physics Department, Royal Institute of Technology, Frescativagen 24, Stockholm 10405, Sweden (telephone: 468161000, e-mail: vladimir.peskov @cern.ch).

## I. Introduction

In the last few years there was an "explosion" of inventions of new micropattern gaseous detectors: Microstrip [1], CAT [2], MICROMEGAS [3], GEM[4], Microdot [5], Microgap [6],Well [7], etc. Although many of these detectors are very different geometrically, one can notice, however, that conceptually they are rather similar [8]: in all cases the gas amplification appears in a small region of strong and focused electrical field. Since this feature is common for all micropattern detectors, it will be interesting to consider the limits of these new micropattern detectors. Results of earlier studies can be found in refs. [9-14], while in this paper we present our latest results and conclusions.

## II. Experimental set up

A detailed description of the experimental setup can be found elsewhere [9,10]. The most important "basic" micro-pattern gaseous detectors were studied: micro-dot [15], micro-gap (obtained from [16]), "compteur a trous" (CAT) [12,15], GEM, MICROMEGAS and thin-gap (0.1 μm) resistive-plate chamber (RPC). The cathode of the RPC (3×3cm$^2$) was done from aluminium and the anode from Pestov glass with resistivity ~$10^{10}$ Ω/□. The inner surface of the anode was covered with chromium strips 10 μm wide, 30 μm pitch.

In some measurements the inner surface of the RPC cathode was covered with a CsI secondary-electron emitter – Fig 1 (see [17] for more details). Two types of emitter structures were used: uniform (300 nm thick) and porous (~1 μm thick). We also studied some combinations of these detectors with preamplification structures: GEM or parallel-plate avalanche chamber (PPAC) [11]. Tests were done in various mixtures of noble gases with quenchers at pressures of 0.05-5 atm. For position resolution studies we used a well-collimated (30 μm) x-ray beam (6-35 keV).

Some tests, in particular with thin gap RPC, were also done in T9 and T10 pion-beam test facilities at CERN [17].

## III. Results.

1. Gain limit at low rate.

We found that the maximum achievable gain ($A$) in micropattern detectors is limited by:

$$A £ Q_{max}/n_0 \qquad (1)$$

where $n_0$ is the number of primary electrons and $Q_{max}$ is the maximum achievable charge in the avalanche, depending on detector design and gas composition and pressure. Typically at 1 atm $Qmax\sim 10^7$ electrons, but each type of micropattern detector has it own limit, which drops rapidly, almost inversely proportional, with pressure.

With two or more multiplication steps $Q_{max}$ increases and may reach ~$10^8$ electrons.

The value of $Q_{max} \sim 10^7$ electrons is always valid for large values of $n_0$ (alpha particles, for instance), but for smaller values of $n_0$ (1-100) $Q_{max}$ could be smaller as well. This is because in order to meet the limit (1) at small $n_0$ values one has to operate the detector at high gains and therefore at high voltages as well. At such voltages breakdown may occur due to construction defects, which are rather common in micropattern detectors. At small $n_0$ one can reach $Q_{max} \geq 10^7$ electrons only with two or more of preamplification steps.

## 2. Gain limit at high rates.

We found that for all micropattern detectors tested, the maximum achievable gain always drops with rate, in some cases almost inversely proportional to it (see Fig.1 in [12]). The same feature appears in some other gaseous detectors, for example in single wire counters [18] or in PPAC's [19], so it seems to be a universal feature of fast detectors [13]. These observations suggest that there is a common physical mechanism, limiting the detectors performance at high rates.

## 3. Position resolution limit

It is clear that in general the position resolution limit in micropattern detectors is determined by the path length of the electrons created by the incident radiation in the gas (primary electrons), by the diffusion in the drift region and by the geometry of the amplification region. Obviously, in order to achieve the best possible position resolution one has to optimize each of these parameters. The minimum possible electron path length can be achieved by using solid converters of radiation [20, 21]. In this case the primary electrons are created mostly inside a thin solid layer and some (sometimes all) primary electrons can be extracted (by applying a strong electric field on the cathode surface) into the gas media and multiplied there. The minimum diffusion spread can be achieved by using the minimum possible gap between the converter and the amplification region. The minimum avalanche size can ensured by using the thinnest possible multiplication gap. These considerations lead to a concept of a micropattern detector with a solid converter immediately followed (without any drift) by an amplification region. One possible design for such a detector is shown in Fig. 1

Fig. 2a shows an image of a 30 µm wide slit placed in front of the detector, perpendicularly to the anode strips and to the electrodes plates. Figures 3b and 3c show the images of the same slit shifted each time by 15 µm in the direction perpendicular to the strips. From the image contrast (ratio of counts from neighbouring strips) one can conclude that a position resolution better than 30 µm was achieved. Note that in a parallel-plate amplification structure, whose maximum multiplication is obtained only for electrons created near the cathode, almost the same position resolution could be achieved without a converter, being however the efficiency very low.

**IV. Discussion**

1 Gain limit

One can see from Fig. 3 that the gain vs. rate curve has different slopes, which we believe, reflect different physical mechanisms, responsible for breakdown. In the following paragraphs we will present possible explanations.

*1.a Very Low rate (<1Hz/mm$^2$)*

One can safely assume that at a low rates breakdown occurs when the space charge field becomes comparable with the applied field (see [22] for more details).

In general, the space charge effect depends on the applied field and on factors that affect the dimensions of the avalanche: pressure, length of the drift region, gap size and gas composition. This is probably the reason why each micropatern detector has its own limit and why it drops rapidly with pressure. As it was shown in a previous work with preamplification structures, $Q_{max}$ increases due to the diffusion [10].

*1.b Medium and high rates (10 to 10$^6$Hz/mm$^2$)*

In previous studies, at least two mechanisms were identified that may contribute to the breakdown of micropattern detectors at medium and high rates: cathode "excitation" [13] and spots with high electric field near the cathodes ("hot" spots [23]). More experiments are needed to fully clarify the main mechanism.

As it was mentioned above, the gain drop with rate is valid for detectors having no "hot" spots, for example single wire counters or PPAC's. One may therefore speculate that cathode excitation dominates in these cases. Breakdown then appears due to jets of electrons, photon or ion feedback loops or their combination [12,13].

*1.c Very high rates (>10$^7$Hz/mm$^2$)*

At very high rates (>10$^7$Hz/mm$^2$), additionally to the cathode excitation effect, plasma-type effects may appear as well. They include the modification of electrical field in the cathode region due to the steady space charge, multistep ionization, gas heating effect and accumulation of excited atoms and molecules [24]. As it was shown in other studies, these mechanisms may create instability leading to breakdown [18].

2 Position resolution limit

The use of solid converters places a real physical limit on the position resolution of micropattern detectors. However, despite the already very satisfactory result obtained, the simple counting method used in this work is far from optimum and the ultimate value of the position resolution should be measured by applying a centre of gravity method. An increase of the gas pressure would further improve the position resolution.

## VI. Conclusions.

As any other instrument, micropattern detectors have some limitations. We believe that identification and understanding of these limits will be useful for those who consider their various applications. The lack of understanding of such limits has led, in the past, to micropattern detectors being proposed for applications that demanded sometimes unrealistic performances.

## VII. Acknowledgements

We thank M. Lemonnier, F. Sauli, L. Ropelewski and Y. Giomataris for useful discussions.

**Figure captions:**

Fig 1: Schematic picture of a thin gap RPC with a CsI converter

Fig. 2: Number of counts measured in a thin-gap RPC with solid converter irradiated by x-rays through a 30 μm slit: a) slit is placed in front of the strip #12; b) slit is placed between the strips #12 and #13; c) slit is placed in front of the strip #13.

Fig. 3: Typical dependence of the maximum achievable total charge in avalanches vs. rate. 1- micropattern detectors. 2-micropattern detectors with preamplification structure.

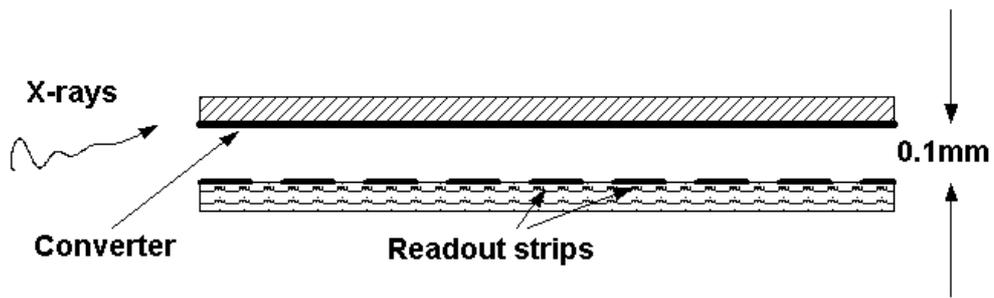

Fig 1: Schematic picture of a thin gap RPC with a CsI converter

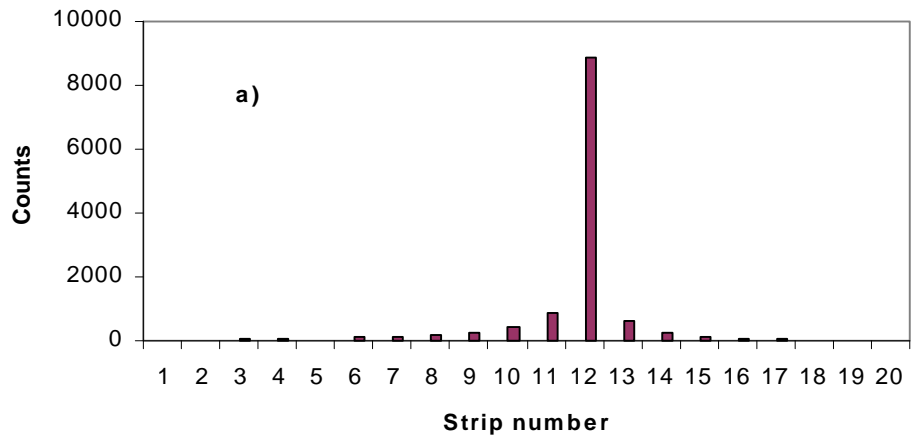

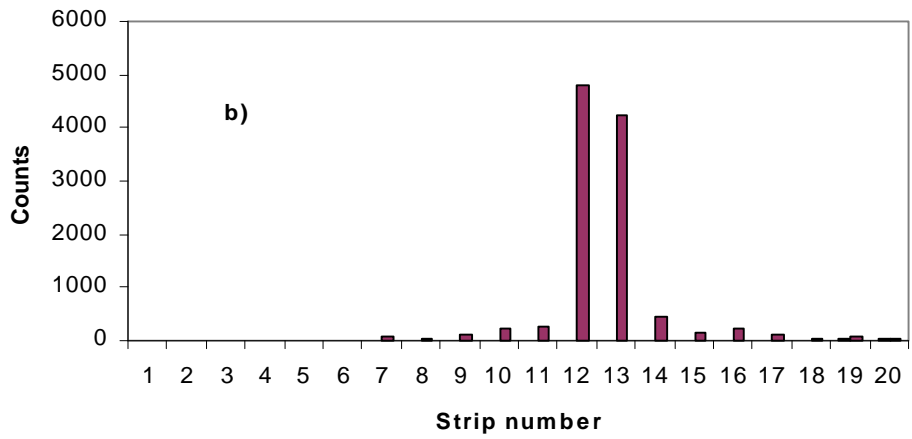

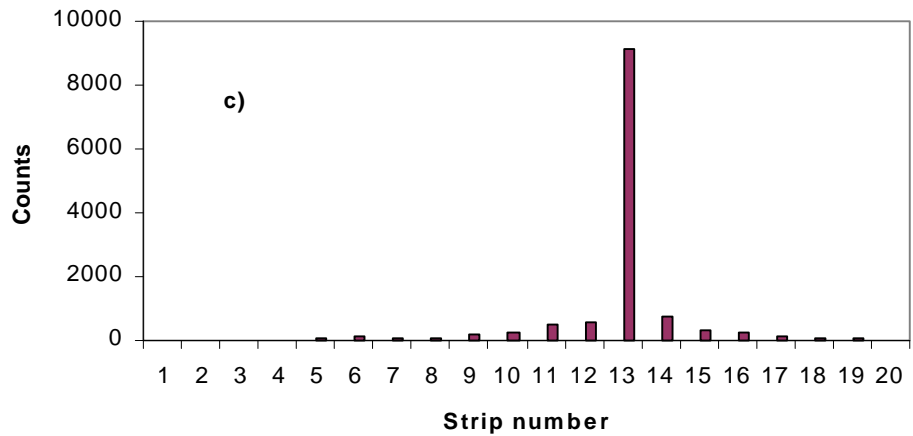

Fig. 2: Number of counts measured in a thin-gap RPC with solid converter irradiated by x-rays through a 30 μm slit: a) slit is placed in front of the strip #12; b) slit is placed between the strips #12 and #13; c) slit is placed in front of the strip #13.

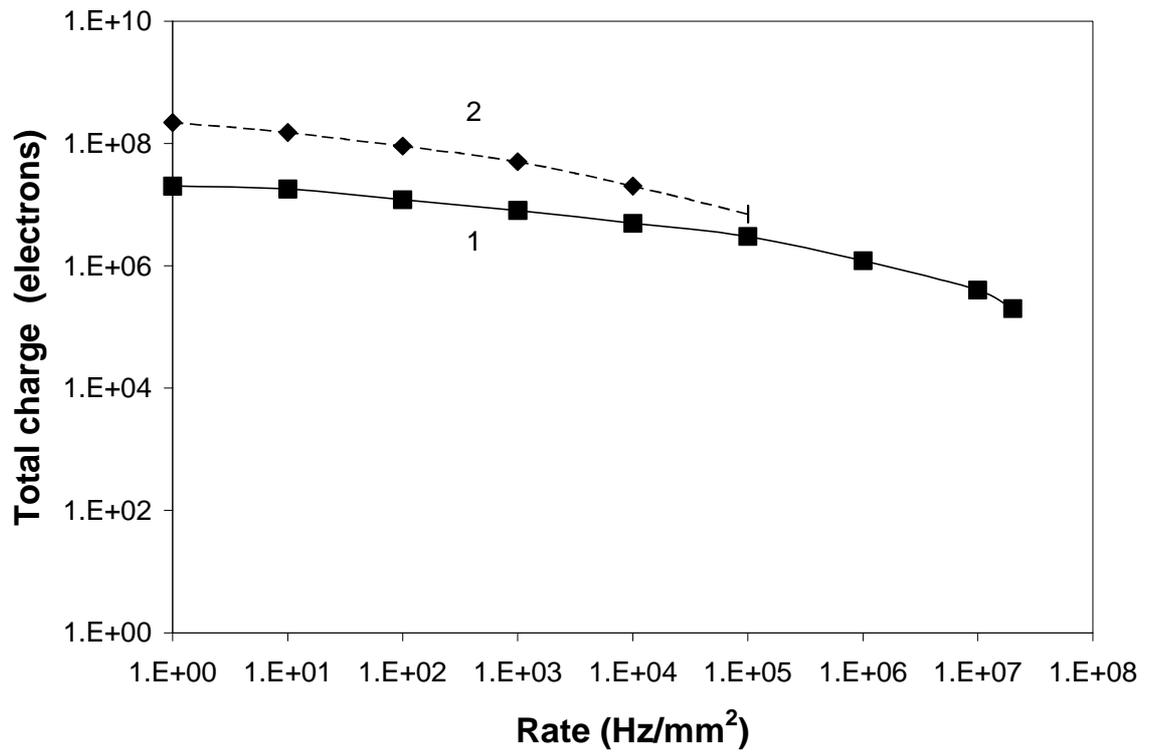

Fig. 3: Typical dependence of the maximum achievable total charge in avalanches vs. rate. 1- micropattern detectors. 2-micropattern detectors with preamplification structure.